\documentclass[prl,twocolumn,aps,showpacs,floats,superscriptaddress]{revtex4}
\usepackage{graphicx}
\def\beq{\begin{equation}}
\def\eeq{\end{equation}}
\def\be{\begin{equation}}
\def\ee{\end{equation}}

\def\iomn{i\omega_n}

\def\t{\mbox{tr}\,}
\def\cG0{{\cal G}_0}
\def\cG{{\cal G}}

\def\bra{\langle}
\def\ket{\rangle}

\def\vk{{\bf k}}

\def\vr{{\bf r}}

\def\vR{{\bf R}}

\def\a{\alpha}

\def\e{\varepsilon}

\def\g{\gamma}

\def\l{\lambda}

\def\hn{\hat{n}}

\def\DS{\Delta S}

\def\DE{\Delta E}
\def\Dv{\Delta V}

\def\dS{\Delta\Sigma\,}

\begin{document}

\title{The $\a-\g$ transition of Cerium is entropy-driven}
\author{ B. Amadon}
\affiliation{CEA,
D{\'e}partement de Physique Th{\'e}orique et Appliqu{\'e}e,
BP 12, 91680 Bruy{\`e}res-le-Ch{\^a}tel, France}
\author{ S. Biermann}
\affiliation{Ecole Polytechnique, Centre de Physique Th{\'e}orique,
91128 Palaiseau Cedex, France}
\author{ A. Georges}
\affiliation{Ecole Polytechnique, Centre de Physique Th{\'e}orique,
91128 Palaiseau Cedex, France}
\author{ F. Aryasetiawan}
\affiliation{Research Institute for Computational Sciences, AIST,
1-1-1 Umezono, Tsukuba Central 2, Ibaraki 305-8568, Japan}
\affiliation{CREST, Japan Science and Technology Agency}

\begin{abstract}
We emphasize, on the basis of experimental data and theoretical calculations,
that the entropic stabilization of the $\g$-phase is the main driving force
of the $\alpha$-$\gamma$ transition of cerium in a wide temperature range
below the critical point. Using a formulation of the total energy as a functional
of the local density and of the f-orbital local Green's functions, we perform
dynamical mean-field theory calculations within a new implementation based on the
multiple LMTO method, which allows to include semi-core states.
Our results are consistent with the experimental energy differences and with the
qualitative picture of an entropy-driven transition, while also
confirming the appearance of a stabilization energy of the
$\a$ phase as the quasiparticle Kondo resonance develops.
\end{abstract}


\maketitle

The $\a-\g$ phase transition of cerium~\cite{Ce,mcmahan_collapse_review}
is a first-order isostructural transition, ending at a second-order critical point at
$T_c\simeq 600{\rm K}$.
When temperature decreases below $T_c$, the volume change
between the two phases increases~\cite{Schiwek,Beecroft}, reaching
15\% at room temperature.
The magnetic susceptibility follows Curie-Weiss behaviour in the
(larger volume) $\gamma$- phase, and is Pauli-like in the
(smaller-volume) $\alpha$- phase.
This is interpreted as
4f electrons being localized in the $\gamma$- phase, giving rise to
local moments and contributing weakly to the electronic bonding
(hence the larger volume). In contrast, in the $\alpha$- phase, the
4f electrons participate in both the bonding and the formation of
quasiparticles.

The detailed mechanism underlying the transition
has been the subject of  debate.
In the Mott transition picture~\cite{johansson}, the focus is put on the
4f orbitals only, while the Kondo volume collapse (KVC) picture~\cite{KVC} emphasizes the
key role of the hybridisation between the 4f electrons and
(spd-) conduction electron states which form broader bands.
In this picture, the stronger hybridisation of the low-volume
$\a$-phase leads to a high Kondo temperature and in turn to a screening
of the 4f local moment, while the high-volume $\gamma$-phase
has a low Kondo temperature, leading in practice to unscreened
moments for $T>T_K^\gamma$.
Photoemission experiments~\cite{photoemission} demonstrate that both phases display
Hubbard bands and hence are strongly correlated. In addition, a
quasiparticle peak is seen in the $\a$ phase only. These observations
are compatible with both pictures. However, a recent theoretical calculation~\cite{optics_haule}
of the optical spectrum, in connection with the experimental
results of Ref.~\cite{optics_vdeb}, has emphasized the importance of hybridisation effects,
in qualitative agreement with the KVC picture.
In both the Mott and KVC pictures, the $\a$- phase is stabilized by energetic effects
(the f-electron kinetic energy in the Mott picture, or the Kondo screening
energy in the KVC picture), while the $\gamma$ phase is stabilized by its
large spin-fluctuation entropy.

In  this letter, we argue that
entropic effects actually play the dominant role in the
transition, at least in the temperature range
$300 {\rm K}<T<T_c$. Using available experimental data,
we estimate the jump in entropy and internal energy
($\DS= S_\g-S_\a$\,,\,$\DE=E_\g-E_\a$) and find that
$T\DS$ is always significantly larger than $\DE$ in
this temperature range.
The second purpose of this article is to examine whether
this conclusion is consistent with total energy calculations
within the LDA+DMFT scheme, a combination of density-functional theory (DFT) within
the local density approximation (LDA) with dynamical mean-field theory (DMFT).
Recently, cerium has been the focus of pioneering
theoretical work~\cite{CeMcMahan,cedmft1,optics_haule,Sakai}
using the LDA+DMFT approach. In Refs.~\cite{CeMcMahan},
the total energy was studied and it was concluded that a negative curvature effect
is apparent already at elevated temperatures ($T\sim 1600 {\rm K}$),
corresponding to the energetic stabilization of the $\a$-phase which
was viewed as ultimately driving the transition.
Here, we reconsider this issue within a new
implementation of LDA+DMFT using the multiple LMTO scheme,
and basing our total-energy calculations on a functional of
the local density and f-orbital Green's function.
Using extensive Quantum Monte-Carlo calculations, we are
able to reach temperatures lower than the experimental $T_c$.
Our results are consistent with the qualitative
picture of an entropy-driven transition, and with the experimentally
measured energy differences.

\begin{figure}
\centering
{\raisebox{0.0cm}{\resizebox{7.0cm}{!}
{\rotatebox{0}{\includegraphics{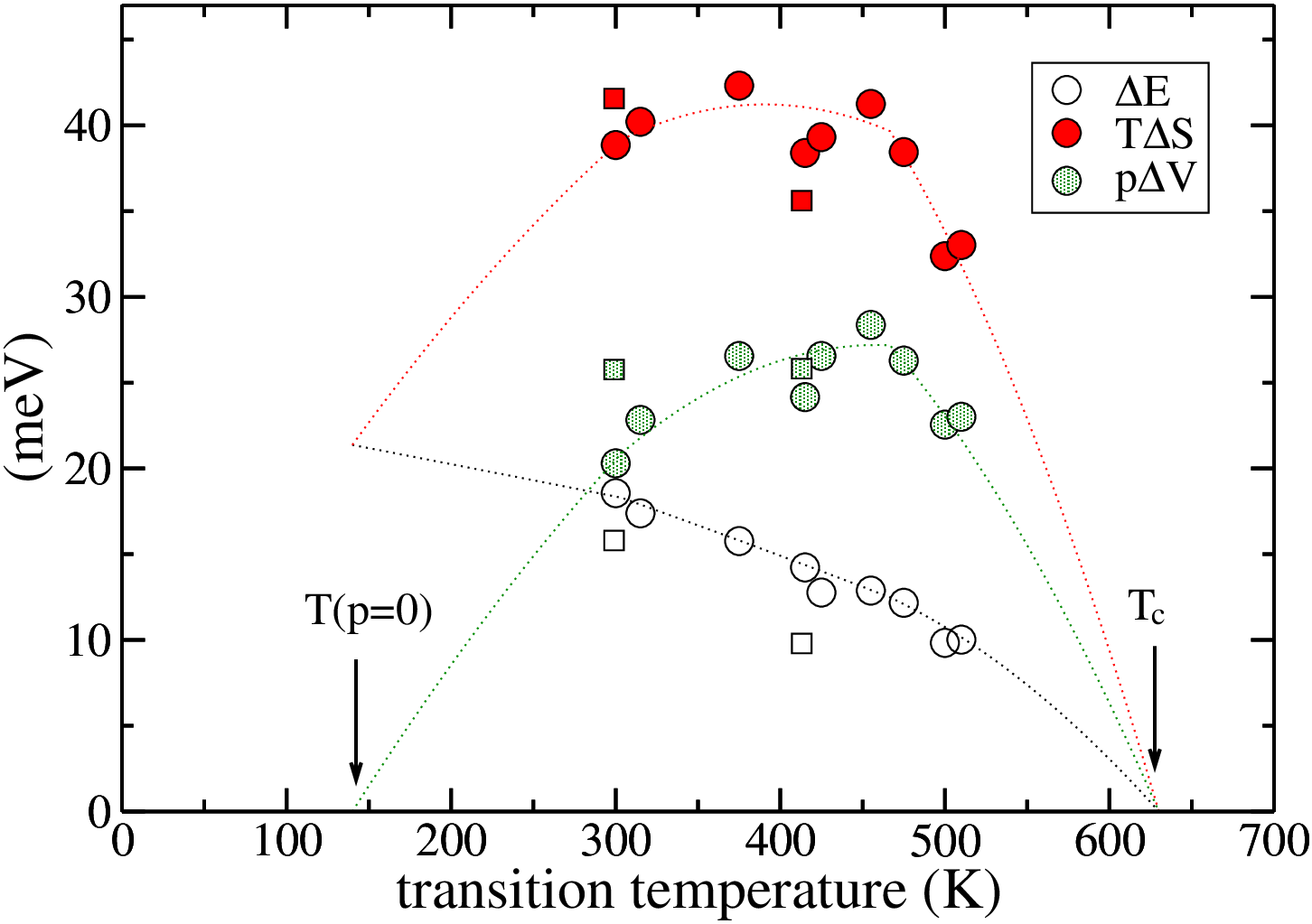}}}}}
\caption{Experimental variation of the entropy term $T\DS$, energy $\DE$
and $p\Delta V$ (obtained as described in the text
from the data of \cite{Beecroft} -circles- and \cite{Schiwek} -squares-),
between the $\g$ and the $\a$ phase
across the transition line in the P-T phase diagram.
Dotted lines are extrapolations based
on exact limits (at T$_{\rm c}$ all three terms vanish,
and $T\DS=\DE$ at $p=0$).
}
\label{fig:exp}
\end{figure}

The Clausius-Clapeyron relation $dp/dT=\Delta S/\Dv$
relates the slope of the transition line to the jump of the
entropy and unit-cell volume $\Dv=V_\g-V_\a$ at the transition.
Furthermore, the continuity of the Gibbs free-energy yields the relation:
$\DE-T\DS+p\Dv=0$. Using available
experimental data~\cite{Beecroft, Schiwek} on $dp/dT$ and $\Dv$, one can
thus determine the three quantities $\DE$, $T\DS$ and $p\Dv$, which
are plotted on Fig.~\ref{fig:exp} as a function of temperature.
As clear from this graph, the entropic term $T\DS$ above room-temperature is
of order $30-40 {\rm meV}$, more than twice as large as the energy difference
$\DE$ between the two phases (of order $10-20 {\rm meV})$.
The energetic stabilization of the $\a$-phase results in
$\DE=E_\g-E_\a>0$, but the difference in free-energy has the
opposite sign: $\Delta F=F_\g-F_a=\DE-T\DS=-p\Dv < 0$ precisely because the
difference of entropy dominates over the energy difference.
We conclude from this analysis that entropic effects
are essential to the physics of the $\a$-$\g$ transition, at
least at room temperature and above.
We note that the relative importance of spin and lattice contributions to the
entropic stabilization of the $\g$-phase
is currently under debate,
with very different conclusions reached from experimental studies of pure
cerium~\cite{Jeong} and of the Ce$_{0.9}$Th$_{0.1}$ alloy~\cite{Manley}.

An accurate calculation of the electronic and lattice free-energy of cerium, from
first principles, is a major challenge.
While the calculation of the entropy is beyond the scope of this article, we
focus in the following on the calculation of the
total energy, at temperatures below the experimental $T_c$, within the
LDA+DMFT framework.
This raises two important methodological questions.
The first one is the proper choice of the valence
states to be included in the starting LDA- hamiltonian.
Indeed, it is mandatory to include semicore states
(in particular $5p$ states) in the valence when computing
the energy, because these states contribute significantly
to the variation of the energy upon compression.
On the other hand, we have found that, within an LMTO-ASA framework,
it is crucial to include the 6p orbitals in the valence in order to
obtain a proper band-structure.
Since in standard implementations of the LMTO-ASA method,
the simultaneous inclusion of $5p$ and $6p$ orbitals
in the valence states is not possible, previous works
using DMFT for Ce were either restricted to spectral
properties only~\cite{cedmft1} or
have treated the different terms in the expression of the total energy
within different 
implementation of DFT
~\cite{CeMcMahan}.
Secondly, in view of the small energy differences
between the two phases (on the scale of $10$ to $20 {\rm meV}$), a precise formulation
of the total energy functional must be used.

The starting Hamiltonian is constructed from an LDA calculation
within the orthogonalized localized basis set of the multiple
LMTO-ASA scheme~\cite{multiplelmto},
retaining $5s$,$5p$,$6s$,$6p$,$5d$ and $4f$ states in the valence
band. We neglect spin-orbit coupling which has little effect on LDA energies
in Cerium.  Many body terms acting on the f-orbitals
are added to this hamiltonian, as well as a double-counting
correction term (as in LDA+U schemes~\cite{anisimov93}), so that the
many-body hamiltonian reads $H=H_{\rm KS} +H_{\rm U} -H_{\rm DC}$ with:
\begin{eqnarray}\label{eq:ham}
H_{\rm KS}=&\sum_{\vk L L'} h^{{\rm KS}}_{LL'}(\vk) c^\dagger_{\vk L}c_{\vk L'}\\
H_U=&\frac{1}{2}\,U\,\sum_\vR\sum_{ab\sigma} \hn_{\vR a}\hn_{\vR b}
\nonumber
\end{eqnarray}
In this expression, $h^{{\rm KS}}_{LL'}(\vk)$ denotes the Kohn-Sham (LDA)
hamiltonian at a given $\vk$-point, expressed in an (orthogonalized) 
LMTO basis set $\chi_L^\vk$,
with $L=\{lm\sigma\}$ and $L'$ running over the full
valence set. The Hubbard term is written in real space, with $\vR$ denoting
atomic positions and $a,b$ running only over the $f$-orbitals.
We use the value of the Coulomb interaction U= 6 eV, computed by constrained
LDA calculations in \cite{cedmft1,mcmahan_collapse_review}.

In order to derive an expression for the total energy, we start from the
(``spectral density'') free-energy functional introduced by Kotliar and
Savrasov~\cite{savrasov_kotliar_spectral_long}, which depends on
the total local electron density $\rho(\vr)$
and the on-site Green's function in the correlated subset of orbitals:
$G_{ab}^{\vR\vR}$ (denoted below $G_{ab}$ for simplicity).
The functional is
constructed by introducing source terms,
$v_{KS}(\vr)-v_c(\vr)$ (the difference of the Kohn Sham
potential $v_{KS}$ and the crystal potential $v_c$),
and $\Delta\Sigma_{ab}(\iomn)$,
coupling to the density operators $\psi^\dagger(\vr)\psi(\vr)$ and to
$\sum_{\vR}\chi^*_a(\vr-\vR)\psi(\vr,\tau)\psi^\dagger(\vr',\tau')\chi_b(\vr'-\vR)
=c_{a\vR}(\tau)c^\dagger_{b\vR}(\tau')$, respectively.
The Luttinger-Ward~\cite{luttinger_ward_2_1960} part of the functional is approximated by
that of the on-site
local many-body hamiltonian $H_U-H_{{\rm DC}}$.
This yields:
\begin{eqnarray}\nonumber
\lefteqn{\Omega[\rho(\vr),G_{ab};v_{KS}(\vr),\dS_{ab}]_{{\rm LDA+DMFT}}}\\ \nonumber
&= &-\frac{1}{\beta}\t\ln[\iomn+\mu+\frac{1}{2}\nabla^2-v_{KS}(\vr)-\chi^*.\dS.\chi]\\ \nonumber
&-&\int d\vr\,(v_{KS}-v_c)\rho(\vr) -\t [G.\dS]  \\ \nonumber
&+&\frac{1}{2}\int d\vr\,d\vr' \rho(\vr) U(\vr-\vr') \rho(\vr')
+ E_{xc}[\rho(\vr)]\\
&+&\sum_\vR\left(\Phi_{{\rm imp}}[G^{\vR\vR}_{ab}]-\Phi_{{\rm DC}}[G^{\vR\vR}_{ab}]\right)
\label{eq:functional}
\end{eqnarray}
Minimization with respect to the sources
gives a functional of the local Green function and the density only.
Stationarity of this functional with respect to $\rho(\vr)$ and $G_{ab}$ yields
the basic equations of LDA+DMFT~\cite{Revueldadmft}, and in particular,
the self-consistency condition for the
local Green's function:
$G_{ab}(\iomn)=\sum_\vk G(\vk,\iomn)_{ab}$. The full Green's function reads:
$\hat{G}^{-1}(\vk,\iomn) =(\iomn+\mu)\cdot 1 -\hat{h}^{{\rm KS}}
+\hat{V}_{DC}-\hat{\Sigma}_{{\rm imp}}(\iomn)$,
with $\Sigma_{{\rm imp}}^{ab}=\delta\Phi_{{\rm imp}}/\delta G_{ab}$ the
local impurity self-energy and $V_{{\rm DC}}^{ab}=\delta\Phi_{{\rm DC}}/\delta G_{ab}$.
From (\ref{eq:functional}), an expression of the total energy within LDA$+$DMFT
can finally be obtained as:
\begin{equation}
E = E_{{\rm DFT}}-\sum_\lambda \e^{{\rm KS}}_\l
+\bra H_{{\rm KS}}\ket+\bra H_U \ket-E_{{\rm DC}}
\label{eq:energy_lda+dmft_1}
\end{equation}
Note that, importantly, the total energy does not simply reduce
to the expectation value $\bra H\ket$ of the many-body
hamiltonian (\ref{eq:ham}).
In (\ref{eq:energy_lda+dmft_1}), $E_{\rm DFT}$ is the expression of the energy
within density-functional theory,
$\sum_\lambda \e^{{\rm KS}}_\l$ is the sum of the occupied Kohn-Sham eigenvalues and
$\bra H_{{\rm KS}}\ket =\t[H_{KS}\hat{G}]$.
Note that these last two terms do not cancel each other, since $\t[H_{KS}\hat{G}]$
is evaluated with the full Green's function including the self-energy, while
$\sum_\l\epsilon^{KS}_\l=\t [H_{KS}\hat{G}_{KS}]$.
Eq.~(\ref{eq:energy_lda+dmft_1}) expresses that the latter term has to be removed from
$E_{DFT}$, in order to correctly take into account the change 
of occupation of the Kohn Sham orbitals.
The LDA+DMFT scheme should in principle be performed
by imposing self-consistency not only on the DMFT
quantities but also on the local density~\cite{PuNature}
(or equivalently on the LDA Hamiltonian), in such a way that
$\rho(\vr)=\bra\vr|\hat{G}|\vr\ket$, i.e including the correlation-induced changes
to the local density.
However, for simplicity and in order to compare to previous
works~\cite{CeMcMahan},
we present as a first step in this paper calculations without
full self-consistency on $\rho(\vr)$. The
double-counting correction term is written in terms of the
LDA occupancy of the f-orbital, as:
$E_{DC}=UN^{\rm f}_{\rm lda}(N^{\rm f}_{\rm lda}-1)/2$.
In order to solve the DMFT equations, we have used the Hirsch-Fye Quantum Monte Carlo
algorithm, and studied the temperature range from 400K to 1600K.
The number of sweeps was adjusted in order to obtain a precision on the
energy of order $20{\rm meV}$, a rather demanding requirement at the
lowest temperature.
The kinetic energy $\bra H_{KS}\ket=\sum_{n,\vk}H_{KS}(\vk)G(\vk,i\omega_n)$
is computed in a direct manner, while the correlation energy $\bra H_U\ket$
is computed from the double occupancy.

\begin{figure}[h]
\begin{tabular}{cc}
{\resizebox{4.0cm}{!}
{\rotatebox{0}{\includegraphics{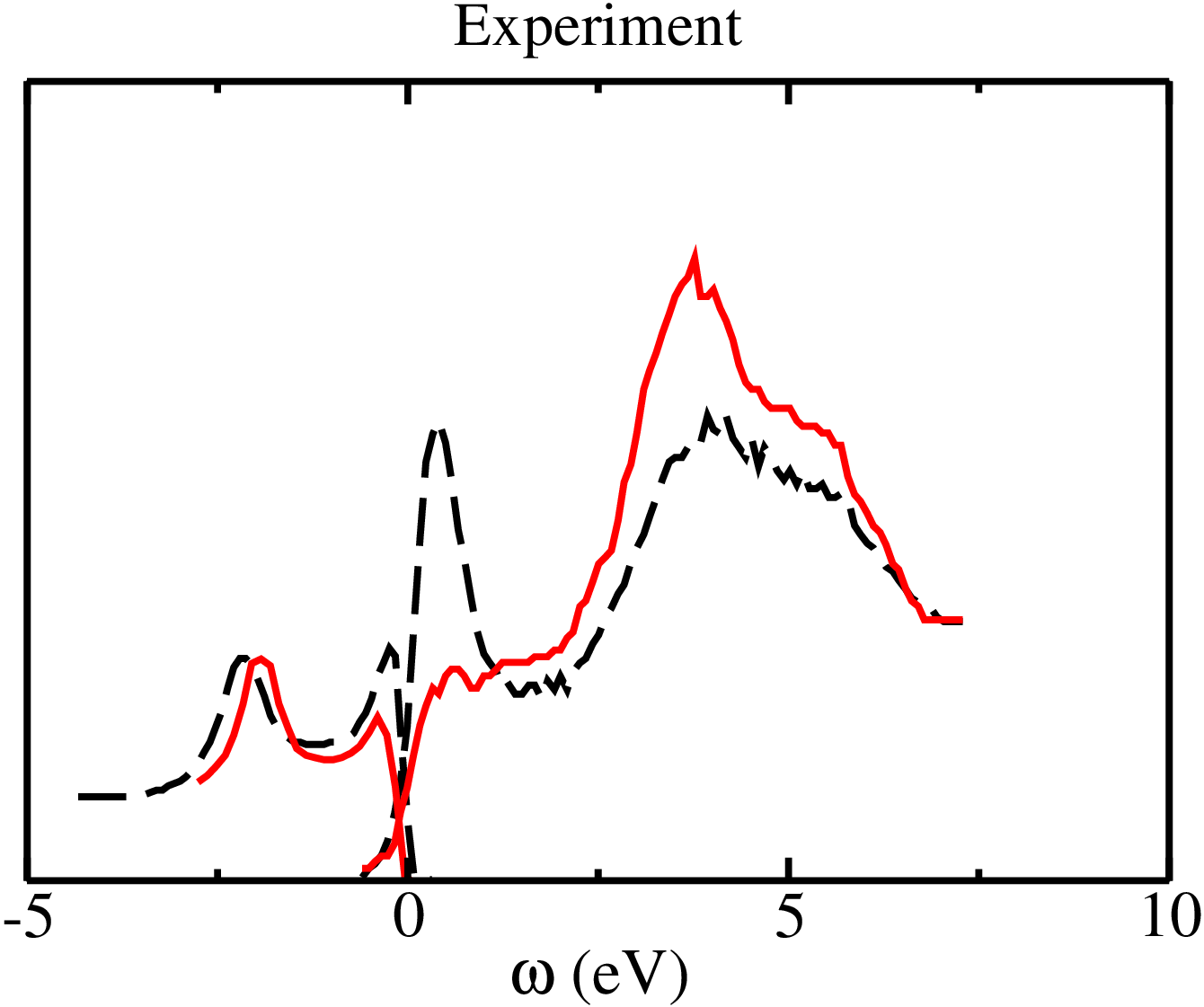}}}} &
{\resizebox{4.0cm}{!}
{\rotatebox{0}{\includegraphics{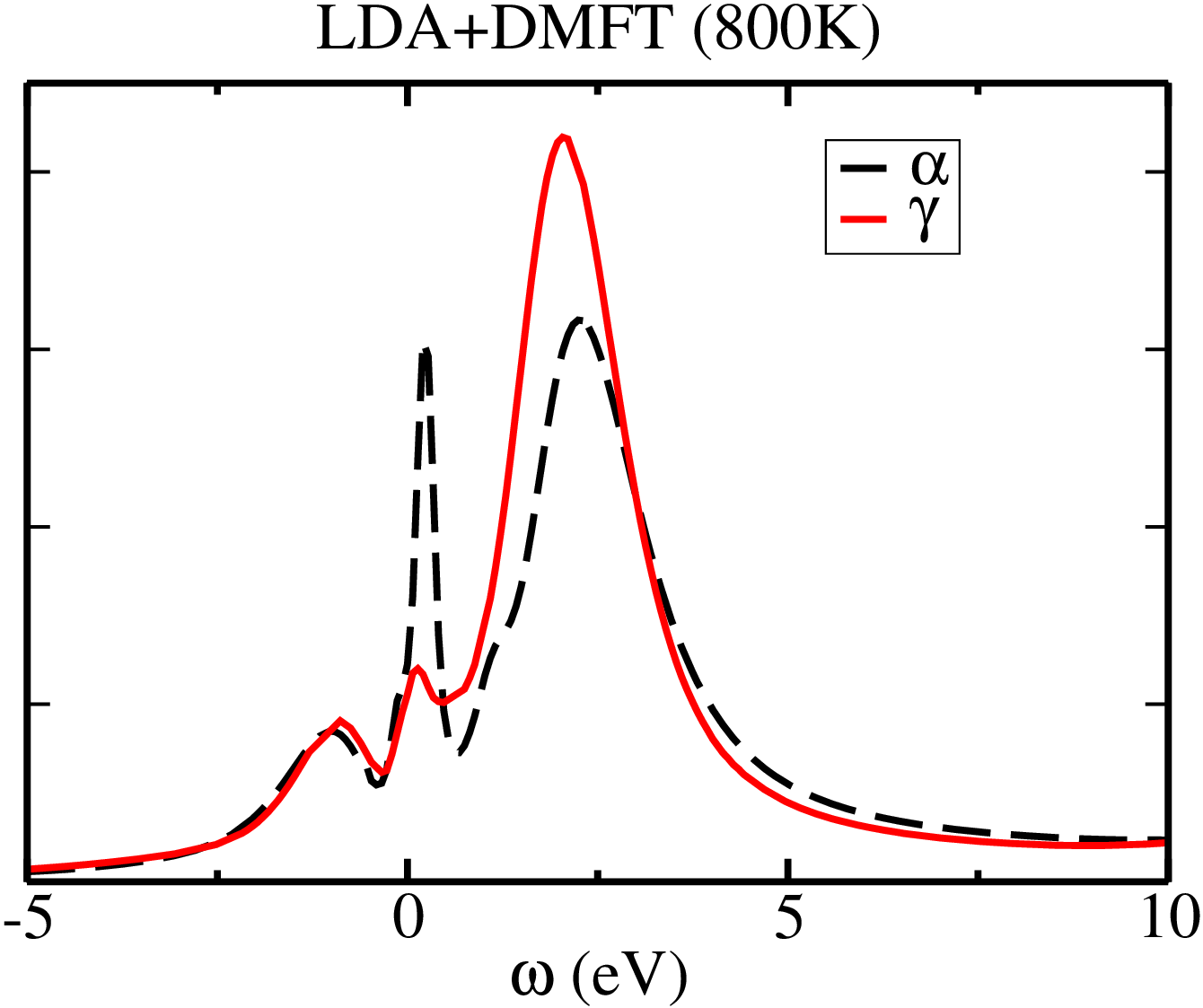}}}} \\
\end{tabular}
\caption{Experimental~\cite{photoemission} and LDA+DMFT theoretical results for the
PES and BIS spectra of $\a$ and $\g$ cerium}
\label{fig:spectres}
\end{figure}
First, we display on Fig.~\ref{fig:spectres} the spectral functions (obtained by maximum-entropy
continuation of our QMC data) for the $\a$ and the $\g$ phase, in comparison
to experimental spectra. As in previous LDA+DMFT studies~\cite{CeMcMahan,cedmft1}, the quasiparticle
peak is correctly described in the $\a$ phase, while Hubbard bands are present
both  in the $\a$ and the $\g$ phases. Their intensities are correct
(although their positions are not very accurately reproduced). These
results give us confidence that the main physical features of both phases are correctly
captured by our calculations.

\begin{figure}[h]
\centering
{\raisebox{0.0cm}{\resizebox{8.4cm}{!}
{\rotatebox{0}{\includegraphics{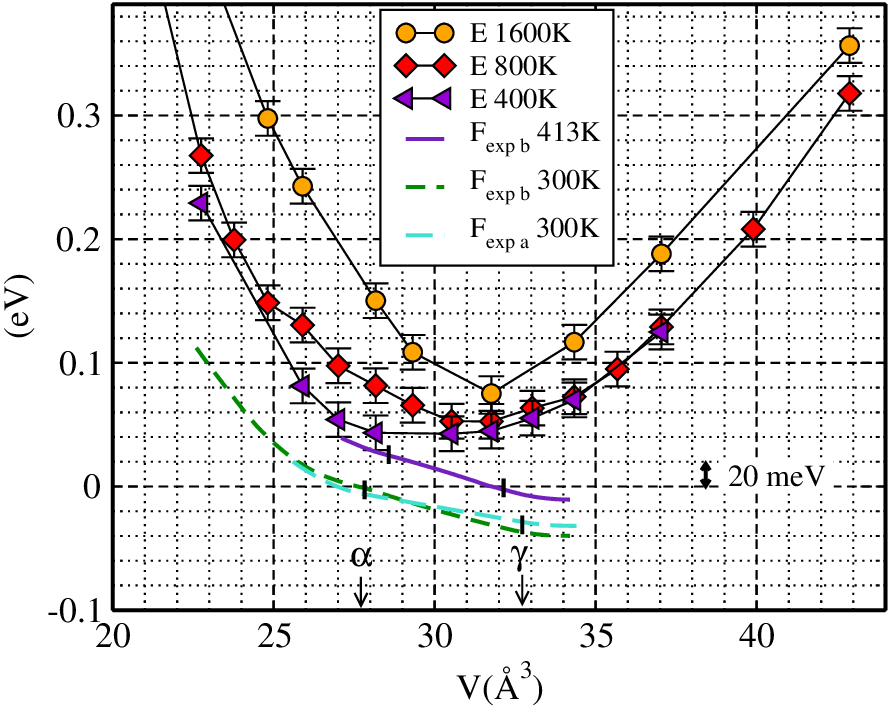}}}}}
\caption{Symbols: internal energy (E) vs. volume curves for cerium, 
computed within LDA+DMFT
for different temperatures. Also shown is the free energy (F), 
calculated from the experimental
pressure vs. volume curves (a:\cite{Jeong}, b:\cite{Schiwek}). 
The position of these experimental curves with respect to 
each other is arbitrary.
Short vertical lines on these curves show the experimental volumes at 
each temperature.
Arrows indicate the volume of each phase at room 
temperature, at the transition pressure.}
\label{fig:evqmc}
\end{figure}

On Fig.~\ref{fig:evqmc}, we display our results for the energy
as a function of volume for three different temperatures
(1600K, 800K and 400K). Statistical error bars of the QMC calculation
are indicated on these plots.
At 1600K we observe a smooth curve with a minimum located at
31.0$\pm 0.5$ {\AA}$^3$.
At 800K, the minimum is located at a somewhat lower volume and we note that
the curvature decreases near the minimum of the curve and in particular
for lower volumes. For 400K, this effect is
strong enough to shift the minimum
to 29$\pm 1$ {\AA}$^3$.
According to experimental results~\cite{Schiwek,Beecroft},
the volume of the $\a$ and $\g$ phases at $400 {\rm K}$ are
28.5$\pm$0.1 and 32.0$\pm$0.1 {\AA}$^3$, and the difference of energy
between these two phases is $13.5\pm 4 {\rm meV}$ (see Fig.~\ref{fig:exp}).
This value is quite consistent with our calculations, even though
a precise theoretical value would require to reduce
the statistical error bars even more.
Overall, we do not find evidence for a region of negative
curvature in the energy versus volume curve. However, 
because $400 {\rm K}$ is below the critical point, a double
tangent should be present in the {\it free-energy} vs. volume $F(V)$.
We have plotted in Fig.~\ref{fig:evqmc} the experimental
free-energy vs. volume curves deduced
from recent pressure versus volume measurements at 413 K~\cite{Schiwek}
and 300K~\cite{Schiwek,Jeong}, by integrating the equation of state:
$F(V)-F(V_0)=-\int_{V_0}^V p(V')dV'$.
For volumes between the equilibrium volumes of the $\a$ and $\g$ phases,
(indicated by short vertical lines for each temperature),
$F(V)$ is taken to be the common tangent.
Comparison of the theoretical energy to the experimental free-energy
suggests that the entropic stabilization of the
$\g$-phase is mainly responsible for the appearance of
a region of negative curvature 
in the {\it free-energy} $F(V)$. Moreover the entropic stabilization $T\DS$ is
of order $40 {\rm meV}$ at $400 {\rm K}$, as seen
from Figs.~\ref{fig:exp} and \ref{eq:energy_lda+dmft_1},
much larger than $\DE$.

\begin{figure}[h]
\centering
{\raisebox{0.0cm}{\resizebox{7.5cm}{!}
{\rotatebox{0}{\includegraphics{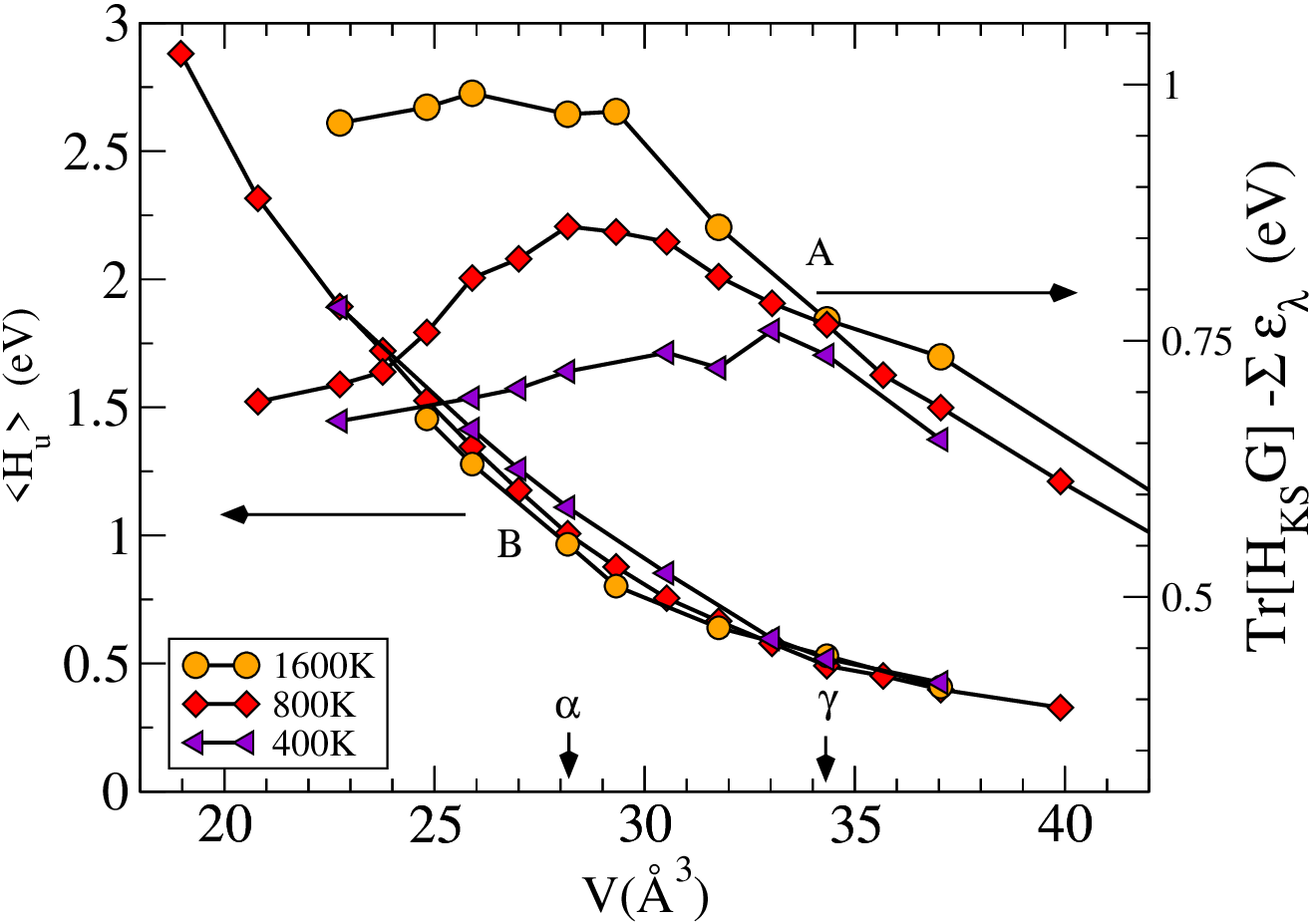}}}}}
\caption{Evolution of  $A=\t[H_{\rm KS}G]-\Sigma \epsilon_\lambda^{\rm LDA}$  and $B=\bra H_{\rm U}\ket$
(note that, in this case, the scale is four time larger) as a function
of volume, computed in DMFT  for different temperatures.}
\label{fig:evdecomp}
\end{figure}

Although we do not find a double tangent in the energy versus volume curve, we do observe a decrease of the curvature and the flattening of the volume dependance of the energy, as temperature is reduced. In order to understand its 
physical origin, we have plotted in Fig.~\ref{fig:evdecomp}, as a function
of volume, the two contributions
$A=\bra H_{\rm KS} \ket -\Sigma_{\lambda} \epsilon^{\rm LDA}_{\lambda}$
(ie the correlation induced changes to the kinetic and hybridization energies)
and $B=\bra H_{\rm U}\ket$ (the interaction energy among $f$ orbitals).
A negative curvature is clearly seen to develop in A as T is reduced, 
consistent with the observed development of the Kondo resonance
(Fig.~\ref{fig:spectres}) and with the stabilization energy of the
$\alpha$ phase as previously emphasized in reference \cite{CeMcMahan}.

\begin{figure}[h]
\centering
{\raisebox{0.0cm}{\resizebox{6.5cm}{!}
{\rotatebox{0}{\includegraphics{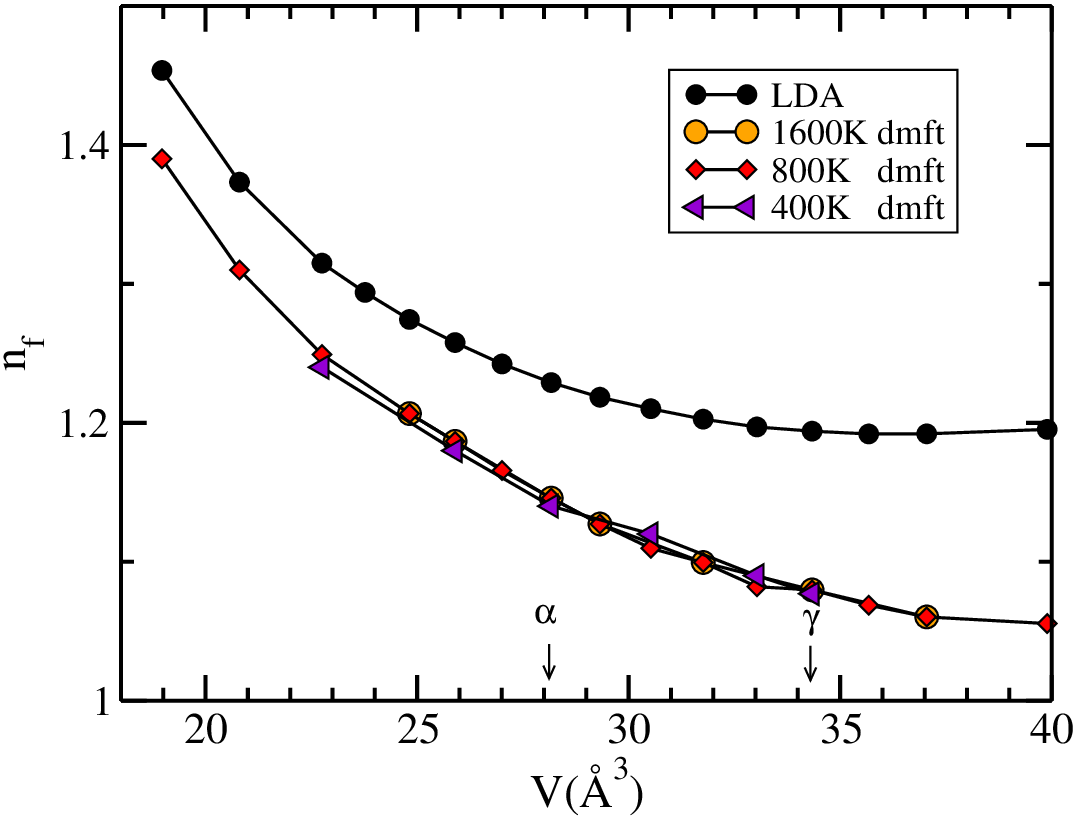}}}}}
\caption{Evolution of the number of f electrons in LDA and DMFT for different
temperatures as a function of volume.}
\label{fig:nfevol}
\end{figure}

Finally, the number of $f$ electrons computed in DMFT is plotted in 
Fig.~\ref{fig:nfevol} as a function of volume. As expected, this number is
very close to 1 in the localized $\gamma$ phase while it increases
at lower volume due to hybridization effects. In contrast to reference 
\cite{CeMcMahan}, we find a monotonous decrease of $n_f$ as 
volume  is increased. One should keep in mind however that
$n_f$ depends on the basis set and the functional.

In conclusion,
we have revisited the problem of the volume-collapse transition
of cerium, emphasizing that it is mainly entropy-driven.
We have presented LDA+DMFT calculations of the total energy,
obtained from a functional of the local density and local
Green's function,
within a new implementation based on the multiple LMTO formalism.
This allows us to include semi-core states and to calculate
the Hamiltonian, the energy, as well as spectra within the same
formalism.
We confirm the development of a contribution to the kinetic and
hybridisation energy
stabilizing the $\a$ phase, as temperature is lowered and the Kondo
quasiparticle resonance develops, in qualitative agreement with the
results of Ref.~\cite{CeMcMahan}.
However, we find that the magnitude of this stabilization energy is too
small to induce a pronounced negative curvature in the total energy curve,
and that the transition is actually driven by entropy effects at least above
room temperature. This is consistent with the experimental measurements of Drymiotis
{\it et al} \cite{Drymiotis}.

\acknowledgments
We acknowledge useful discussions with
A.~I. Lichtenstein (who also shared with us his QMC code), as well
as with K.~Haule, K.~Held, G.~Kotliar and A.~K~McMahan.
This work has been supported by CEA, CNRS, Ecole Polytechnique and
by an RTN contract of the E.U (HPRN-CT-2002000295).
F. A. thanks NAREGI Nanoscience Project, MEXT, 
Japan, for financial support.
{\it Note added:} as this paper was being written, we learned of the
preprint cond-mat/0504380 by A.K.McMahan, in which LDA+DMFT
calculations are performed for several rare-earths,
including also spin-orbit coupling. The conclusions of this recent
work regarding cerium are qualitatively similar to those of
Ref.~\cite{CeMcMahan}.

\bibliographystyle{apsrev}

\end{document}